\documentclass[aps,prd,preprint,eqsecnum,tightenlines,floats,
showpacs,superscriptaddress]{revtex4}
\usepackage{amssymb} 
\usepackage{hyperref} 
\usepackage{graphicx}
\usepackage{subfigure}

\def\be{\begin{equation}}
\def\ee{\end{equation}}
\def\beb{\begin{equation*}}
\def\eeb{\end{equation*}}
\def\bea{\begin{eqnarray}}
\def\eea{\end{eqnarray}}
\def\beab{\begin{eqnarray*}}
\def\eeab{\end{eqnarray*}}

\def\bi{\begin{itemize}}
\def\ei{\end{itemize}}
\def\p{\partial}

\def\vp{{{\varphi}}}

\begin{document}

%\preprint{UFIFT-HET-11-*}

\date{April 21, 2015}

\title{Linear Newtonian perturbation theory from the 
Schr\"odinger-Poisson equations}

\author{Nilanjan Banik, Adam J. Christopherson, Pierre Sikivie 
and Elisa Maria Todarello}

\affiliation{Department of Physics, University of Florida, 
Gainesville, FL 32611, USA}

\begin{abstract}

We obtain solutions to the coupled Schr\"odinger-Poisson equations.  The 
solutions describe the evolution of cold dark matter density perturbations 
in an otherwise homogeneous expanding Friedmann universe.  We discuss the 
relationships between descriptions of cold dark matter in terms of a 
pressureless fluid, in terms of a wavefunction, of a classical scalar 
field, and a quantum scalar field.  We identify the regimes where the 
various descriptions coincide and where they differ.

\end{abstract}


\maketitle

\section{Introduction}

The identity of dark matter is among the most tantalizing questions 
in science today \cite{Bertone}.  Fortunately we live in an era where 
a large number of observations directly or indirectly bear upon this 
question.  Foremost among these are measurements of galactic rotation 
curves, observations of gravitational lensing by dark matter clumps on 
various scales, the cosmic microwave background anisotropy observations, 
and broad surveys of the  visible matter distribution such as the Sloan 
Digital Sky Survey.  

The observations require the dark matter to be, in first approximation, 
{\it cold} and {\it collisionless}.  Collisionless means that the main 
force acting on dark matter and therefore the main force by which dark 
matter manifests its presence is gravity.  The dark matter may have 
in addition weak non-gravitational interactions but the observations 
are consistent with the absence of non-gravitational interactions.
That the dark matter is cold means that its primordial velocity 
dispersion is small.  By primordial velocity dispersion we mean 
the velocity dispersion that the dark matter particles have even 
in the absence of density perturbations.  An upper limit of order 
$10^{-8} c$ on the primordial velocity dispersion follows from the 
requirement that free streaming of the dark matter particles does 
not erase density perturbations on the smallest scales on which 
they are observed.

Particle candidates for the dark matter are also meaningfully 
constrained by the requirement that they must fit comfortably 
with the particles that are already known to exist, i.e. those 
described by the Standard Model.  There are three broad categories 
of dark matter candidates that are thought to fit well into the 
existing scheme of particle physics:  weakly interacting massive 
particles (WIMPs),  axions or axion-like particles, and sterile 
neutrinos.  WIMPs are motivated by supersymmetric extensions of 
the Standard Model.  Their mass is typically 100 GeV, and their 
primordial velocity dispersion of order $10^{-12} c$.  Axions 
are motivated by the Peccei-Quinn solution of the strong CP 
problem, the puzzle within the Standard Model why the strong 
interactions are P and CP invariant.  The axion mass is thought 
to be of order $10^{-5}$ eV/$c^2$ and the axion primordial velocity 
dispersion of order $10^{-17} c$.  Sterile neutrinos have mass 
of order a few keV/$c^2$ and primordial velocity dispersion of 
order $10^{-8} c$, at the limit of what is allowed.  For this 
reason, sterile neutrinos are sometimes called ``warm dark 
matter".  Axions and WIMPs are definitely cold dark matter.

Cold collisionless dark matter may be described in the 
linear regime of the evolution of density perturbations 
as a pressureless fluid.  This is the description of cold 
dark matter in calculations of the cosmic microwave 
background anisotropies \cite{Dodelson}.  Since cold 
dark matter plays an important role in this context and 
the calculations agree very well with the observations, 
the pressureless fluid description has high credibility.  
To obtain the cosmic microwave background anisotropies a 
full general relativistic treatment \cite{genrel} is 
necessary because the relevant evolution occurs in part 
on length scales of order the horizon.  However, on length 
scales much less than the horizon (i.e. for wavevectors 
much larger than the Hubble rate) dark matter density perturbations 
are correctly described by Newtonian gravity.  Linear Newtonian 
perturbation theory \cite{Peebles} is simple, well understood 
and agrees with the general relativistic description on length 
scales much less than the horizon, where many of the interesting 
phenomena in large scale structure formation occur.  It is therefore 
a very useful tool.

L. Widrow and N. Kaiser \cite{Widrow} pointed out that, on scales 
much less than the horizon, cold collisionless dark matter can be 
described by a wavefunction satisfying the Schr\"odinger-Poisson 
equations.  As is discussed below, in Section III, the wavefunction 
description is in many ways more powerful than the pressureless 
fluid description.  It allows the introduction of velocity dispersion 
whereas the fluid description allows none.  It can be used to describe 
multi-streaming and caustics in the non-linear regime, whereas the 
fluid description breaks down in that regime.  Indeed, Widrow and 
Kaiser carried out numerical simulations of structure formation 
using a wavefunction satisfying the Schr\"odinger equation, in lieu 
of $N$ bodies satisfying Newton's force law equation.  Several such 
simulations have since been carried out \cite{Schrod}.
 
In the present paper, we reproduce the results of Newtonian 
linear perturbation theory using a wavefunction solving the 
Schr\"odinger-Poisson (sometimes called Schr\"odinger-Newton) 
equations.  As far as we are aware, this had not been done before 
although related work, also using the Schr\"odinger equation to 
analyze the growth of density perturbations in the early universe, 
can be found in Refs. \cite{Moroz,Szapudi, Johnston}. One of our 
motivations is to show that the formalism does indeed work as 
expected.  However, our main motivation is to prepare the ground 
for an in-depth study of the dynamical evolution of axion dark 
matter.

The proposal that the dark matter may be axions originates with the 
papers of Ref.\cite{axdm} which showed that axions are copiously 
produced during the QCD phase transition, at a temperature of order 
1 GeV.  The estimate of the cosmological energy density of the axions 
thus produced was obtained by a simple classical treatment of the 
axion field.  The axions are extremely weakly interacting and therefore 
collisionless.  They are non-relativistic shortly after being produced 
and subsequently red-shifted by the expansion of the universe.  Thus 
they are very cold, as was already mentioned.  It was emphasized in 
Ref.~\cite{Ipser} that the axions produced during the QCD phase 
transition behave as cold dark matter on all scales much longer 
than the wavelength of the axion field, hereafter called the de 
Broglie wavelength.  So, although axions are much lighter than WIMPs, 
they behave in many circumstances in the same way as WIMPs.  However, 
there are differences.

Obviously, axions behave differently from WIMPs on the length scale 
of their de Broglie wavelength.  One manifestation of this is the 
existence of a Jeans' length for axion dark matter.  This was 
originally pointed out by the authors of Refs.\cite{Khlopov,Bianchi}.  
The formula for the Jeans' length is given in Eq.~(\ref{Jeans}) below. 
For QCD axions, those that solve the strong CP problem and have masses 
conservatively in the range $10^{-2}$ to $10^{-12}$ eV, the Jeans length 
is far too short to affect structure formation in an observable way.  
However, we may also have axion-like particles (ALPs) with much smaller
masses.  Indeed, string theory predicts the existence of numerous axion 
and axion-like fields \cite{Svrcek,Arvanitaki}.  If the ALP mass is 
in the $10^{-21}$ to $10^{-24}$ eV mass range and below, the Jeans' 
length is large enough (kpc and larger) to have observable effects.  
Structure formation on length scales less than the Jeans' length is 
suppressed.  There is a long standing discrepancy by which observations 
show less structure on small scales than is predicted by N-body simulations.  
Many authors have proposed to resolve this by hypothesizing that the dark 
matter is an extremely light scalar field, with mass of order $10^{-21}$ eV 
or less \cite{m21}.

Less obviously, axions differ from WIMPs because they thermalize and 
form a Bose-Einstein condensate (BEC) \cite{CABEC,therm}.  Axions 
thermalize as a result of their gravitational self-interactions when 
the photon temperature is of order 500 eV.  Their thermalization time 
becomes shorter than the age of the universe then.  When they thermalize, 
almost all axions go to the lowest energy state available to them.  In 
this they differ from the other dark matter candidates.  It was shown 
in Ref. \cite{CABEC} that, on all scales of observational interest, 
density perturbations in axion BEC behave in exactly the same way as 
those in ordinary cold dark matter provided the density perturbations 
are within the horizon and in the linear regime.  On the other hand, 
when density perturbations enter the horizon, and in second order of 
perturbation theory, axions generally behave differently from ordinary 
cold dark matter because the axions rethermalize so that the state most 
axions are in tracks the lowest energy available state.  Axion BEC 
explains the occurrence of caustic rings of dark matter in galactic 
halos and their observed radii \cite{case}. It also solves the galactic 
angular momentum problem \cite{Banik}.  The observations require that 
at least 35\% of the dark matter is axions \cite{Banik}.

Systems dominated by gravitational self-interactions are inherently
unstable.  In this regard the axion BEC differs from the BECs that
occur in superfluid $^4$He and dilute gases \cite{Pethik}. The axion 
fluid is subject to the Jeans gravitational instability and this is 
so whether the axion fluid is a BEC or not \cite{NBPS}.  The Jeans
instability causes density perturbations to grow at a rate of order
the Hubble rate $H(t)$, i.e. on a time scale of order the age of the
universe at the moment under consideration.  Each mode of the axion
fluid is Jeans unstable.  However when the thermalization time is 
shorter than the age of the universe, the rate at which quanta of 
the axion field jump between modes is faster than the rate at which 
the Jeans instability develops.  So the modes are essentially frozen 
on the time scale over which the axions thermalize.

Our long term goal is to clarify the dual role of gravity in the 
evolution of the axion BEC.  On the one hand gravity causes Jeans 
instability of the axion field modes.  On the other, gravity causes 
axions to jump between those field modes.  In the present paper, we 
take two steps towards this goal.  In Section II we solve the 
Schr\"odinger-Poisson equations for self-gravitating collisionless 
dark matter.  Our solutions describe the homogeneous expanding 
Friedmann universe and density perturbations therein.  They 
also provide a complete set of states for the axions to occupy.
In Section III, we discuss the various relationships between 
descriptions of cold dark matter in terms of a pressureless fluid, 
in terms of a wavefunction, of a classical scalar field, and a 
quantum scalar field, identifying the regimes where the various 
descriptions coincide and where they differ.  In Section IV, we 
summarize our conclusions.

\section{Wavefunction description of linear perturbations}

Consider a fluid composed of a huge number $N$ of particles that
are all in the same quantum-mechanical state.  The wavefunction
$\psi(\vec{r},t)$ for the state satisfies the Schr\"odinger
equation
\be
i\p_t\psi(\vec{r},t)=\Big(-\frac{1}{2m}\nabla^2
+m\Psi(\vec{r},t)\Big)\psi(\vec{r},t)\,,
\label{Schro}
\ee
where $m$ is the particle mass and $\Psi(\vec{r},t)$ is the
Newtonian gravitational potential. In this section, we set
$\hbar = c = 1$. The density of particles in the fluid is
\be
n(\vec{r},t)= N \psi^*(\vec{r},t)\psi(\vec{r},t)~~\ .  
\label{den}
\ee
Let us assume that the only kind of matter present is the 
fluid of particles.  The gravitational potential is then 
given by the Poisson equation
\be \nabla^2\Psi=4\pi Gm n(\vec{r},t)~~\ .
\label{Pois}
\ee
The fluid density satisfies the continuity equation
\be
\partial_t n + \vec{\nabla}\cdot\vec{j} = 0~~\ ,
\label{cont}
\ee
where
\be
\vec{j} = {N \over 2mi}(\psi^* \vec{\nabla} \psi
- \psi \vec{\nabla} \psi^*)~~\ .
\ee
The fluid velocity $\vec{v}(\vec{r},t)$ is defined by  
$\vec{j}(\vec{r},t) \equiv n(\vec{r},t) \vec{v}(\vec{r},t)$.
If we write 
$\psi(\vec{r},t) = \sqrt{n(\vec{r},t)} e^{i \beta(\vec{r},t)}$,
then
\be
\vec{v} = {1 \over m} \vec{\nabla} \beta~~\ .
\label{velo}
\ee
The velocity field satisfies the Euler-like equation      
\be
\partial_t \vec{v} + (\vec{v}\cdot\vec{\nabla}) \vec{v} =
- \vec{\nabla} \Psi  - \vec{\nabla} q~~\ ,
\label{Euler}
\ee
where
\be
q = - {1 \over 2 m^2} {\nabla^2 \sqrt{n} \over \sqrt{n}}~~\ .
\ee          
$q$ is commonly referred to as ``quantum pressure".  Eqs.~(\ref{cont})
and (\ref{Euler}) follow from Eq.~(\ref{Schro}).

We want to use Eqs.~(\ref{Schro}) and (\ref{Pois}) to describe the
evolution of density perturbations in an otherwise homogeneous
Friedmann universe.  The universe may be open or closed, or in
between.   However, because our description uses Newtonian gravity,
the cosmological constant is set equal to zero.

The wavefuction describing the homogeneous universe is
\be         
\psi_0(\vec{r},t)=\sqrt{n_0(t)}{\rm e}^{i\frac{1}{2}mH(t)r^2}~~\ ,
\label{homog}
\ee
where $H(t)$ is the Lema\^itre-Hubble expansion rate.
Indeed Eqs.~(\ref{velo}) and (\ref{homog}) imply
\be
\vec{v} = H \vec{r}~~\ .
\ee  
The imaginary part of the Schr\"odinger equation 
is satisfied provided
\be
\partial_t n_0 + 3H n_0=0\,,\\
\ee
and its real part is satisfied provided
\be 
\Psi_0 = - {1 \over 2}(\partial_t H + H^2) r^2~~\ .
\ee
The Poisson equation then implies the acceleration equation
\be
\partial_t H+H^2=-\frac{4\pi G}{3}mn_0(t)~~\ .
\ee
The continuity and acceleration equations may be combined as
usual to yield the Friedmann equation
\be
H(t)^2 + {K \over a(t)^2} = {8 \pi G \over 3} m n_0(t)~~\ ,
\ee
where $K = -1,0, +1$ depending on whether the universe is open, 
critical or closed, and $a(t)$ is the scale factor defined by 
$H(t) = {\dot{a} \over a}$.

We now consider perturbations about this background:
\be
\psi(\vec{r},t)= \psi_0(\vec{r},t)+\psi_1(\vec{r},t)\,.
\ee
The perturbation is Fourier transformed in terms of
comoving wavevector $\vec{k}$ as follows:
\be
\psi_1(\vec{r},t)=
\psi_0(\vec{r},t)\int d^3{k}\,\,\psi_1(\vec{k},t)                 
{\rm e}^{i\frac{\vec{k}\cdot\vec{r}}{a(t)}}~~\ .
\ee
Likewise the perturbation to the gravitational potential
\be
\Psi_1(\vec{r},t)=\int d^3{k}\,\, \Psi_1(\vec{k},t)
{\rm e}^{i\frac{\vec{k}\cdot\vec{r}}{a(t)}}\,.
\ee
The Schr\"odinger-Poisson equations expanded to linear
order in the perturbations imply
\be
\label{eq1}
i\partial_t\psi_1=-\frac{1}{2m}\nabla^2\psi_1
+m(\Psi_0\psi_1+\Psi_1\psi_0)\,,
\ee 
and
\be
\label{eq2}
\nabla^2\Psi_1=4\pi G m(\psi_0^*\psi_1+\psi_0\psi_1^*)\,.
\ee
It is useful to introduce the functions
\bea               
\delta(\vec{k},t)&\equiv& \psi_1(\vec{k},t)+\psi_1^*(-\vec{k},t)\,,\\
\eta(\vec{k},t)&\equiv& \psi_1(\vec{k},t)-\psi_1^*(-\vec{k},t)\,,
\eea
in terms of which Eqs.~(\ref{eq1}) and (\ref{eq2}) become      
\bea
\label{eq:rev}
i\partial_t \delta(\vec{k},t)&-&\frac{k^2}{2ma^2(t)}\eta(\vec{k},t)=0\,,\\
\label{eq:sev}
i\partial_t \eta(\vec{k},t)&+&\Bigg(\frac{8\pi G m^2 n_0(t)}{k^2}a^2(t)
-\frac{k^2}{2ma^2(t)}\Bigg)\delta(\vec{k},t)=0\,.           
\eea
These can be combined into one, second order differential
equation for $\delta(\vec{k},t)$:              
\be
\partial_t^2 \delta(\vec{k},t) + 
2 H(t) \partial_t \delta(\vec{k},t)
-4\pi G\rho \delta(\vec{k},t)+\frac{k^4}{4m^2a^4(t)}\delta(\vec{k},t)=0~~\ ,
\label{denev}
\ee
where $\rho = m n_0$.  The Fourier components of the perturbation to the 
wavefunction are given by
\be
\psi_1(\vec{k},t) = {1 \over 2} \delta(\vec{k},t) +
i {m a(t)^2 \over k^2} \partial_t \delta(\vec{k},t)~~ \,
\ee
in terms of the solutions to Eq.~(\ref{denev}).

The perturbation to the number density is               
\be
n_1(\vec{r},t)=\left|\psi_0(\vec{r},t)\right|^2\int d^3k\,\,
\Big(\psi_1(\vec{k},t)+\psi_1^*(-\vec{k},t)\Big)
{\rm e}^{i\frac{\vec{k}\cdot\vec{r}}{a(t)}}\,,
\ee
and so the density contrast is
\be
\delta(\vec{r},t)=\frac{n_1(\vec{r},t)}{n_0(\vec{r},t)}
=\int d^3k\,\,\delta(\vec{k},t){\rm e}^{i\frac{\vec{k}\cdot\vec{r}}{a(t)}}~~\ .
\ee
By expanding
\be
\psi(\vec{r},t)=\sqrt{n_0(t)+n_1(\vec{r},t)}
{\rm e}^{i\Big(\beta_0(\vec{r},t)+\beta_1(\vec{r},t)\Big)}\,,
\ee
one finds that              
\be
\beta_1(\vec{r},t)= \frac{1}{2i}\int d^3k\,\, \eta(\vec{k},t)
{\rm e}^{i\frac{\vec{k}\cdot\vec{r}}{a(t)}}~~\ .
\ee
Hence Eq.~(\ref{eq:rev}) implies 
\be           
\vec{v}_1({k},t)=\frac{ia(t)\vec{k}}{\vec{k}\cdot\vec{k}}            
\partial_t\delta(\vec{k},t)~~\ ,                         
\ee        
which is the same relationship between the velocity perturbation 
and the density contrast as in the standard description of cold 
dark matter in terms of a pressureless fluid.  Eq.~(\ref{denev}) 
is also the standard second order differential equation governing 
the evolution of the density contrast, except for the last term.  
It arises due to quantum pressure in Eq.~(\ref{Euler}) and produces 
a Jeans length \cite{Khlopov,Bianchi}
\be
\ell_J = (16 \pi G \rho m^2)^{-{1 \over 4}} = 
1.01 \cdot 10^{14} {\rm cm} 
\left({10^{-5} {\rm eV} \over m}\right)^{1 \over 2}
\left({10^{-29} {\rm g/cm}^3 \over \rho}\right)^{1 \over 4}~~\ .
\label{Jeans}
\ee
For $k > {a(t) \over \ell_J}$,  the Fourier components of the density 
perturbations oscillate in time.  For $k << {a(t) \over \ell_J}$, 
the most general solution of Eq.~(\ref{denev}) is
\be                        
\delta(\vec{k},t)=A(\vec{k})\left(\frac{t}{t_0}\right)^{2/3}+B(\vec{k})
\left(\frac{t_0}{t}\right)~~\ ,
\ee
in the critical universe case [$a(t) \propto t^{2 \over 3}$]. $A(\vec{k})$ 
and $B(\vec{k})$ are the amplitudes of growing and decaying modes, 
respectively.  On distance scales much larger than the Jeans length, 
the wavefunction description coincides in all respects with the fluid 
description.

Let us mention briefly that the wavefunction can also describe 
rotational modes, provided vortices are introduced.  See, for 
example ref. \cite{Banik}.  In a region where 
$\vec{\nabla} \times \vec{v} \neq 0$, the vortices have the direction 
of $\vec{\nabla} \times \vec{v}$ and have density (number of vortices 
per unit area) ${m \over 2 \pi}|\vec{\nabla} \times \vec{v}|$.  By 
Kelvin's theorem, the vortices must move with the fluid. Therefore, 
in an expanding universe, the density of vortices decreases as $a(t)^{-2}$.  
Hence $\vec{v} \propto a(t)^{-1}$ for rotational modes, which is again the 
usual result.

\section{Discussion}

We saw in the previous section that density perturbations in the 
early universe may be described by a wavefunction which solves the
Schr\"odinger-Poisson equations and that on length scales large 
compared to the Jeans length, Eq.~(\ref{Jeans}), the resulting 
description coincides with that in terms of a pressureless fluid.  
It is our purpose in the present section to place this result in 
a wider physical context.

First let us state that, although it appears that the wavefunction
description had not been explicitly given before, it is no surprise
that it exists since the Schr\"odinger equation implies the continuity 
equation and the Euler-like equation (\ref{Euler}).  These two equations
are the basic equations describing a fluid.  The only difference is the 
quantum pressure term in Eq.~(\ref{Euler}) but that term is unimportant 
on distance scales large compared to the de Broglie wavelength.  The 
Jeans length of Eq.~(\ref{Jeans}) can be viewed as the de Broglie 
wavelength of the minimum energy state in a region of density $\rho$.  
Indeed in such a region, the gravitational potential is 
$\Psi = {2 \pi \over 3} G \rho r^2$ and hence the energy 
of a trial wavefunction of width $b$ is of order 
\be
E(b) \sim {1 \over 2 m b^2} + {2 \pi \over 3} G \rho m b^2~~~\ .
\ee
$E(b)$ reaches its minimum for 
$b \sim ({4 \pi \over 3} G \rho m^2)^{-{1 \over 4}} \sim \ell_J$.

However the mathematical equivalence of the two descriptions
hides important physical differences.  This is perhaps best 
illustrated by an example. Consider the wavefunction
\be
\psi(\vec{r},t)= A \Big({\rm e}^{i \vec{k}\cdot\vec{r}}+
{\rm e}^{-i \vec{k}\cdot\vec{r}}\Big){\rm e}^{-i\omega t}\,.
\label{example}
\ee
where $A$ is a constant and $\omega = {\vec{k}\cdot\vec{k} \over 2m}$.  
It solves the Schr\"odinger equation for a free particle.  The fluid 
with $N$ particles in the state of wavefunction $\psi(\vec{x},t)$ has 
two flows, both with density $n_1=n_2= N |A|^2$, and with velocities
$\vec{v}_1=\frac{\vec{k}}{m}$ and $\vec{v}_2 = - \frac{\vec{k}}{m}$.  On 
the other hand, Eqs.~(\ref{den}) and (\ref{velo}) map $\psi(\vec{r},t)$ onto 
a fluid whose density is $n(\vec{r}) = 4 N |A|^2 \cos^2(\vec{k}\cdot\vec{r})$ 
and whose velocity $\vec{v} = 0$.  The two descriptions are mathematically 
equivalent in the sense that $n(\vec{r},t)$ and $\vec{v}(\vec{r},t)$ 
satisfy  Eqs.~(\ref{den}) and (\ref{Euler}) because $\psi(\vec{r},t)$
satisfies Eq.~(\ref{Schro}).  But the two descriptions are not 
physically equivalent.  They are physically equivalent only if
we average over distances large compared to the wavelength 
${2 \pi \over k}$ and if we ignore the velocity dispersion 
$\Delta v = {k \over m}$.  The spatial averaging is justified 
in the limit $k \rightarrow \infty$.  Ignoring the velocity 
dispersion is justified in the limit, ${k \over m} \rightarrow 0$.  
The two limits are compatible only if $m \rightarrow \infty$.  
This indicates that the physical differences between the 
wavefunction and fluid description disappear completely only 
in the limit where the dark matter particle is very heavy.

The fluid description never allows velocity dispersion since 
the velocity field $\vec{v}(\vec{r},t)$ has a single value at 
every point.  In contrast, the wavefunction description allows 
velocity dispersion and multi-streaming.  The wavefunction 
description is richer therefore.  It can describe everything 
that a fluid describes but the reverse is not true.  Whether 
either description is correct depends on the situation at 
hand.  

Consider a dark matter particle with properties typical 
of a WIMP candidate: $m \sim 100$ GeV, density today 
$n_0 \sim 10^{-8}/{\rm cm}^3$, and primordial velocity 
dispersion today $\delta v_0 \sim 10^{-12}$.  The de Broglie 
wavelength associated with the primordial velocity dispersion 
is of order $10^{-3}$ cm, much smaller than the average interparticle 
distance of order 5 m.  The particles are highly non-degenerate
therefore.  Provided that their primordial velocity dispersion 
is in fact irrelevant to whatever phenomenon is under study (free 
streaming would be an exception since it is a direct result of 
primordial velocity dispersion), the particles can be described 
as a pressureless fluid.  They can also be described by a wavefunction.  
The wavefunction description will in almost all respects be equivalent 
to the pressureless fluid description but, unlike the latter, it 
allows the inclusion of velocity dispersion and its associated 
effects. The wavefunction description is also applicable to the 
non-linear regime, after shell crossing, when the fluid description 
in terms of a single velocity field $\vec{v}$ breaks down.  The 
wavefunction description is more powerful because it packs more 
information.  The wavefunction varies on a length scale of order 
$10^{-3}$ cm in the example given.  The fluid description is far 
coarser.

Next consider a dark matter candidate typical of axions or 
axion-like particles: spin zero, $m \sim 10^{-5}$ eV, density 
today $n_0 \sim 10^9/{\rm cm}^3$, and primordial velocity dispersion 
today $\delta v_0 \sim 10^{-17}$.  The de Broglie wavelength associated 
with the primordial velocity dispersion is of order $10^{18}$ cm.  The 
axion fluid is highly degenerate.  The average occupation number of 
those states that are occupied is huge, of order $10^{61}$.  This 
suggests that axion dark matter is well described by a classical 
scalar field  $\vp(\vec{r},t)$.  The remainder of this section 
considers whether this is so.

A classical scalar field satisfies the Klein-Gordon 
equation
\be
- c^2 D^\mu \partial_\mu \vp + \omega_0^2 \vp + {\lambda \over 3!} \vp^3 = 0
\label{KG}
\ee
where $D_\mu$ is the covariant derivative of general relativity.
We allow the presence of a self-interaction 
${\cal L}_{\vp^4} = - {\lambda \over 4 !} \vp^4$ in the 
action density.  First, let us emphasize that the classical 
field theory, Eq.~(\ref{KG}), has no notion of axion.  The 
axion is the quantum of the quantized scalar field which we 
call $\Phi(\vec{r},t)$.  There is no more notion of axion in 
Eq.~(\ref{KG}) than there is a notion of photon in Maxwell's 
equations.  Also there is no notion of mass since the mass $m$ 
is the energy of an axion, divided by $c^2$.  Henceforth, for 
the sake of clarity, we no longer set $\hbar$ and $c$ equal to 
one.  $\omega_0$ in Eq.~(\ref{KG}) is not the axion mass but 
the oscillation frequency of small perturbations in the classical 
scalar field in the infinite wavelength limit.

In the Newtonian limit of general relativity, the metric is 
$g_{00} = - c^2 - 2 \Psi$, $g_{0i} = 0$, $g_{ij} = \delta_{ij}$.   
Eq.~(\ref{KG}) becomes then:
\be
(\partial_t^2 - c^2 \nabla^2 + \omega_0^2) \vp 
+ {\lambda \over 3 !} \vp^3 
- ({2 \over c^2} \Psi \partial_t^2 + 
\vec{\nabla} \Psi \cdot \vec{\nabla}
+ {1 \over c^2} \partial_t \Psi \partial_t) \vp = 0~~\ .
\label{KGN}
\ee 
We obtain the non-relativistic limit of this equation 
by setting 
\be 
\vp(\vec{r},t) = 
\sqrt{2}~{\rm Re} [{\rm e}^{- i \omega_0 t} \phi(\vec{r},t)]
\label{nonrel}
\ee
and neglecting $\Psi$ versus $c^2$, $\partial_t \phi$ versus 
$\omega_0 \phi$, $\partial_t \Psi$ versus $\omega_0 \Psi$, and 
dropping terms proportional to ${\rm e}^{2 i \omega_0 t}$ and 
${\rm e}^{- 2 i \omega_0 t}$ which indeed oscillate so fast as 
to effectively average to zero.  Eq.~(\ref{KGN}) becomes then 
\be
i \partial_t \phi = - {c^2 \over 2 \omega_0} \nabla^2 \phi
+ {\lambda \over 8 \omega_0} |\phi|^2 \phi + 
{\omega_0 \over c^2} \Psi \phi~~\ .
\label{Schrod2}
\ee
To obtain the Schr\"odinger equation
\be
i \hbar \partial_t \psi = - {\hbar^2 \over 2m} \nabla^2 \psi
+ V(\vec{r},t) \psi~~~\ ,
\label{Schrod3}
\ee
substitute
\be 
\phi(\vec{r},t) = \sqrt{\hbar \over \omega_0} \psi(\vec{r},t)
\ee
in Eq.~(\ref{Schrod2}) and set $m= {\hbar \omega_0 \over c^2}$.  
The potential energy in Eq.~(\ref{Schrod3}) is given, in the 
sense of mean field theory, by 
\be
V(\vec{r},t) = m \Psi(\vec{r},t) + 
{\hbar^4 \lambda \over 8 m^2 c^4} |\psi(\vec{r},t)|^2~~\ .
\label{pot}
\ee
The Newtonian limit of Einstein's equation is the 
Poisson equation, Eq.~(\ref{Pois}).  The non-linear 
version of Schr\"odinger's equation obtained by 
substituting Eq.~(\ref{pot}) into Eq.~(\ref{Schrod3}) 
is commonly called the Gross-Pitaevskii equation.

So the Schr\"odinger equation describes the dynamics of a classical 
scalar field in the non-relativistic limit.  This result is not new 
of course.  We reproduced it here to prepare the ground for the actual 
question we want to discuss, namely whether dark matter axions (or 
axion-like particles) are described by the Schr\"odinger-Poisson 
equations.  Clearly, if axions are described by a classical scalar 
field, the answer is yes as we have just seen.  But the axion is a 
quantum field.  It may behave like a classical field some of the 
time or perhaps even all the time, but this is something that has 
to be proved.  It cannot be merely assumed.

Inside a cubic box of volume $V=L^3$ with periodic boundary 
conditions, the quantum axion field may be expanded (see for 
example Ref. \cite{therm})
\be
\Phi(\vec{r},t) = \sum_{\vec{n}} 
\sqrt{\hbar \over 2 \omega_{\vec n} V}
[a_{\vec{n}}(t) 
{\rm e}^{{i \over \hbar} \vec{p}_{\vec{n}}\cdot\vec{r}} + 
a_{\vec{n}}^\dagger(t) 
{\rm e}^{- {i \over \hbar} \vec{p}_{\vec{n}}\cdot\vec{r}}]~~\ ,
\label{expand}
\ee
where $\vec{n} = (n_1, n_2, n_3)$ with $n_k~(k = 1,2,3)$
integers, $\vec{p}_{\vec n} = {2 \pi \hbar \over L} \vec{n}$,
$\omega = {c \over \hbar} \sqrt{\vec{p} \cdot \vec{p} + c^2 m^2}$.  
The $a_{\vec n}$ and $a_{\vec n}^\dagger$ are annihilation and 
creation operators satisfying canonical equal-time commutation 
relations:
\be
[a_{\vec n}(t), a_{\vec n^\prime}^\dagger (t)] =
\delta_{{\vec n},{\vec n^\prime}}~~,~~
[a_{\vec n}(t), a_{\vec n^\prime}(t)] = 0~~\ .
\label{can}
\ee
The classical field limit is the limit where
$\hbar \rightarrow 0$ with $\hbar {\cal N}$ held fixed, 
where ${\cal N}$ is the quantum occupation number of the 
state described by a particular solution of the classical 
field equations.  The Hamiltonian for the quantum field 
$\Phi(\vec{r},t)$ which satisfies Eqs.~(\ref{Schrod2}) and 
(\ref{Pois}) in the classical field limit is \cite{therm} 
\be
H~=~\sum_{\vec n} \hbar \omega_{\vec n}~a_{\vec n}^\dagger a_{\vec n}
~+ \sum_{\vec{n}_1,\vec{n}_2,\vec{n}_3,\vec{n}_4}
{1 \over 4}~\hbar \Lambda_{\vec{n}_1,\vec{n}_2}^{\vec{n}_3,\vec{n}_4}~
a_{\vec{n}_1}^\dagger a_{\vec{n}_2}^\dagger a_{\vec{n}_3} a_{\vec{n}_4}~~\ ,
\label{axHamil}
\ee
where $\Lambda_{\vec{n}_1,\vec{n}_2}^{\vec{n}_3,\vec{n}_4}$ is the 
sum of two terms:
\be
\Lambda_{\vec{n}_1,\vec{n}_2}^{\vec{n}_3,\vec{n}_4} = 
\Lambda_{s~\vec{n}_1,\vec{n}_2}^{~~\vec{n}_3,\vec{n}_4}~~+~~
\Lambda_{g~\vec{n}_1,\vec{n}_2}^{~~\vec{n}_3,\vec{n}_4}~~\ .
\label{Lam}
\ee
The first term   
\be
\Lambda_{s~\vec{n}_1,\vec{n}_2}^{~~\vec{n}_3,\vec{n}_4}
= + {\lambda \hbar^3 \over 4 m^2 c^4 V}~
\delta_{\vec{n}_1 + \vec{n}_2, \vec{n}_3 + \vec{n}_4}
\label{selfc}
\ee
is due to the $\lambda \Phi^4$ type self-interactions. 
The second term
\be
\Lambda_{g~\vec{n}_1,\vec{n}_2}^{~~\vec{n}_3,\vec{n}_4}
= - {4 \pi G m^2 \hbar \over V}
\delta_{\vec{n}_1 + \vec{n}_2, \vec{n}_3 + \vec{n}_4}~
\left({1 \over |\vec{p}_{\vec{n}_1} - \vec{p}_{\vec{n}_3}|^2}
+ {1 \over |\vec{p}_{\vec{n}_1} - \vec{p}_{\vec{n}_4}|^2}\right)
\label{gravc}
\ee
is due to the gravitational self-interactions.
The Heisenberg equations of motion are 
\be
i \dot{a}_{\vec{n}_1} = - {1 \over \hbar} [H, a_{\vec{n}_1}] = 
\omega_{\vec{n}_1} a_{\vec{n}_1} + 
{1 \over 2} \sum_{\vec{n}_2, \vec{n}_3, \vec{n}_4}
\Lambda_{\vec{n}_1,\vec{n}_2}^{\vec{n}_3,\vec{n}_4}
a_{\vec{n}_2}^\dagger a_{\vec{n}_3} a_{\vec{n}_4}~~\ .
\label{Heom}
\ee
We may likewise expand the classical field 
\be
\vp(\vec{r},t) = \sum_{\vec{n}}
\sqrt{\hbar \over 2 \omega_{\vec n} V}
[A_{\vec{n}}(t) 
{\rm e}^{{i \over \hbar} \vec{p}_{\vec{n}}\cdot\vec{r}} +
A_{\vec{n}}^*(t) 
{\rm e}^{- {i \over \hbar} \vec{p}_{\vec{n}}\cdot\vec{r}}]~~\ .
\label{expandc}
\ee
The Fourier components $A_{\vec{n}}(t)$ satisfy 
\be
i \dot{A}_{\vec{n}_1} = 
\omega_{\vec{n}_1} A_{\vec{n}_1} +
{1 \over 2} \sum_{\vec{n}_2, \vec{n}_3, \vec{n}_4}
\Lambda_{\vec{n}_1,\vec{n}_2}^{\vec{n}_3,\vec{n}_4} 
A_{\vec{n}_2}^* A_{\vec{n}_3} A_{\vec{n}_4}~~\ .   
\label{ceom}
\ee
Eqs.~(\ref{Heom}) and (\ref{ceom}) look similar but, as 
we will see, their physical implications are different 
because the $a_{\vec{n}}(t)$ are operators whereas the 
$A_{\vec{n}}(t)$ are c-numbers.

Let us define the operator 
\be
{\cal N}_{\vec{n}}(t) = a_{\vec{n}}^\dagger(t)~a_{\vec{n}}(t)~~\ ,
\label{occu}
\ee
i.e. the occupation number at time $t$ of the state labeled 
$\vec{n}$.  It was shown in Ref.~\cite{therm} that the Hamiltonian 
of Eq.~(\ref{axHamil}) implies the following operator evolution 
equation
\be
\dot{\cal N}_l = \sum_{k,i,j= 1} {1 \over 2} |\Lambda_{ij}^{kl}|^2
\left[{\cal N}_i {\cal N}_j ({\cal N}_l + 1)({\cal N}_k + 1)
- {\cal N}_l {\cal N}_k ({\cal N}_i + 1)({\cal N}_j + 1)\right]
2 \pi \delta(\omega_i + \omega_j - \omega_k - \omega_l)~~\ . 
\label{Bolq}
\ee
To remove unnecessary clutter, we replaced $\vec{n}_j$ by $j$.
The derivation of Eq.~(\ref{Bolq}) assumes only that the energy 
dispersion $\delta \omega$ of the highly occupied states is much 
larger than the relaxation rate $\Gamma = {1 \over \tau}$.  $\tau$ 
is the relaxation time, defined as the time scale over which the 
distribution $\{{\cal N}_j\}$ changes completely.  When 
$\delta \omega >> \Gamma$, the system is said to be in the ``particle 
kinetic" regime.  The same derivation that yields Eq.~(\ref{Bolq}) but 
applied to the classical counterparts 
\be
N_{\vec{n}}(t) = A_{\vec{n}}^*(t) A_{\vec{n}}(t)
\label{coccu}
\ee
yields 
\be
\dot{N}_l = \sum_{k,i,j= 1} {1 \over 2} |\Lambda_{ij}^{kl}|^2
\left[ N_i N_j N_l + N_i N_j N_k - N_l N_k N_i - N_k N_l N_j\right]
2 \pi \delta(\omega_i + \omega_j - \omega_k - \omega_l)~~\ ,
\label{Bolc}
\ee
again in the particle kinetic regime.  Eqs.~(\ref{Bolq}) and 
(\ref{Bolc}) are clearly different, and they imply different 
outcomes.

Consider the process $i + j \rightarrow k + l$ where two quanta, 
initially in states $i$ and $j$ move to states $k$ and $l$.  Assuming
$\Lambda_{kl}^{ij} \neq 0$, this process always occurs in the quantum 
theory when the initial states are occupied.  In the classical theory, 
the corresponding process occurs only if, in addition, at least one of 
the final states is occupied.  In particular the scattering of two waves 
does not happen in the classical theory of Eq.~(\ref{Schrod2}) if the 
only waves present are the two waves in the initial state.  The quantum 
theory behaves differently because the final state oscillators have 
zero point oscillations.  Incidentally, this observation shows that the 
oft repeated statement that the quantum and classical theories differ 
only by loop effects is incorrect.

After a sufficiently long time, the classical and quantum systems 
thermalize and reach an equilibrium distribution.  The time scale
of thermalization of the classical system is of the same order of 
magnitude as that of the quantum system \cite{therm} but the outcomes 
of thermalization are different.  In the quantum case, Eq.~(\ref{Bolq}) 
implies that the equilibrium distribution $\{{\cal N}_j\}$ is such that 
\be
{\cal N}_i {\cal N}_j ({\cal N}_l + 1)({\cal N}_k + 1)   
- {\cal N}_l {\cal N}_k ({\cal N}_i + 1)({\cal N}_j + 1) = 0
\label{conq}
\ee
for every quartet of states  such that $\omega_i + \omega_j 
= \omega_k + \omega_l$.  Let us call $\epsilon = \hbar \omega$, 
and rewrite Eq.~(\ref{conq}) as  
\be
(1 + {1 \over {\cal N}_i})(1 + {1 \over {\cal N}_j}) = 
(1 + {1 \over {\cal N}_k})(1 + {1 \over {\cal N}_l})
\label{conq2}
\ee
whenever $\epsilon_i + \epsilon_j = \epsilon_k + \epsilon_l$. 
Eq.~(\ref{conq2}) is solved by 
\be
\epsilon_i = C \ln\left[1 + {1 \over {\cal N}_i}\right]
\ee
where $C$ is a constant.  Upon identifying $C = k_B T$, this 
is seen to be the Bose-Einstein distribution
\be
{\cal N}_i = {1 \over {\rm e}^{\epsilon_i \over k_B T} - 1}~~\ .
\label{BE}
\ee
On the other hand Eq.~(\ref{Bolc}) implies that the equilibrium 
distribution $\{N_j\}$ for the classical case satisfies 
\be
(N_i + N_j) N_k N_l = N_i N_j (N_k + N_l) 
\label{conc}
\ee
whenever $\epsilon_i + \epsilon_j = \epsilon_k + \epsilon_l$.
Eq.~(\ref{conc}), which may be rewritten as
\be
{1 \over N_i} + {1 \over N_j} = {1 \over N_k} + {1 \over N_l}~~\ ,
\ee
is solved by 
\be
{\cal N}_i = C {1 \over \epsilon_i}~~ .
\ee
Upon identifying $C = k_B T$, we have
\be 
N_i \epsilon_i = k_B T~~\ ,
\label{Boltz}
\ee 
which is indeed the standard result for classical oscillators
at temperature $T$: each oscillator has energy $k_B T$ on 
average.

We conclude that axion dark matter is not described by a classical 
field when it thermalizes.  Interactions are seen  to have a dual 
role.  They determine the behaviour of the classical field as 
described by Eq.~(\ref{Schrod2}), or Eqs. (\ref{ceom}) and 
(\ref{Bolc}) which follow directly from Eq.~(\ref{Schrod2}).  
Eq.~(\ref{Schrod2}) has a set of solutions which we may label 
$\phi_{\vec{\alpha}}(\vec{r},t)$.  The solutions describe
the states that the axions may occupy in the quantum theory.
In the quantum theory, however, the interactions have the 
additional role of causing transitions between the various 
states $\phi_{\vec{\alpha}}(\vec{r},t)$.  Indeed if there 
were no such transitions the outcomes of thermalization in 
the quantum and classical theories would be the same.  We 
just saw that they are not.

Let $\tau$ be the time scale over which the distribution 
$\{{\cal N}_{\vec{\alpha}}\}$ of the axions, over the states 
described by the classical solutions $\phi_{\vec{\alpha}}(\vec{r},t)$, 
changes completely.  We call $\tau$ the {\it relaxation} or 
{\it thermalization} time scale.  Note that full thermalization 
only happens generally on a time scale much longer than $\tau$. 
(The time scale for ``full" thermalization depends on the degree 
of thermalization required and is therefore less robustly defined 
than $\tau$.)  On time scales short compared to $\tau$, axion dark 
matter behaves as a classical field because only relatively few 
transitions take place between the states described by the classical 
solutions $\phi_{\vec{\alpha}}(\vec{r},t)$.  On time scales long 
compared to $\tau$, the axions are not described by a classical 
field because their distribution over those states changes completely.  
On time scales long compared to $\tau$ dark matter axions form a 
Bose-Einstein condensate (BEC) since they are highly degenerate 
and their number is conserved.  The time scale for BEC formation 
is the relaxation time $\tau$ \cite{Semikoz,therm,Saikawa,Jaeckel}.  
Axion BEC means that almost all axions go the lowest energy state 
available.  The question is then: does $\tau$ ever become shorter 
than the age of the universe $t$ at that moment?  It was found in 
Ref.\cite{CABEC,therm} that $\tau \sim t$ during the QCD phase 
transition at a temperature of order 1 GeV when cold dark matter 
axions are first produced.  The axions thermalize briefly then as 
a result of their $\lambda \Phi^4$ interactions.  Here and elsewhere, 
by the word `thermalize', we mean that the axion distribution relaxes 
and begins to approach a thermal distribution.  We do not mean that 
they thermalize fully.  This brief period of thermalization has no 
known observational consequences.  However, when the photon temperature 
reaches of order 500 eV, cold dark matter axions thermalize anew as 
a result of their gravitational self-interactions and this does have 
observational consequences, as was already mentioned in the Introduction.

\section{Summary}

We derived solutions of the coupled Schr\"odinger and Poisson equations.
The solutions describe the homogeneous expanding matter-dominated 
universe and density perturbations therein.  The description is
identical to that obtained by treating the dark matter as a 
pressureless fluid, on all scales much larger than the de Broglie 
wavelength of the wavefunction.  In a number of respects, the wavefunction 
description is simpler and hence superior.  It has fewer degrees of freedom 
since a wavefunction is two real fields whereas a fluid is described by 
four real fields, the density and the three components of velocity.  The 
mathematics is simpler as well.  Even though it has only two real fields, 
the wavefunction can describe rotational modes.  On the other hand, the 
meaning of the wavefunction is less intuitively obvious.

We considered whether the wavefunction and fluid descriptions are 
equivalent in general.  They are equivalent in the sense that the 
density and velocity fields, given by Eqs.~(\ref{den}) and (\ref{velo}),
satisfy the fluid equations if the wavefunction satisfies the 
Schr\"odinger equation.  However, the wavefunction and the fluid 
describe objects which in general are not physically the same. In 
particular the wavefunction description allows velocity dispersion 
and multi-streaming whereas the fluid description does not.  The 
physical distinction between the two descriptions disappears completely
only in the limit where the mass of the dark matter particle goes to 
infinity.  Because WIMPs are relatively heavy, the wavefunction 
and fluid descriptions of WIMP dark matter are equivalent whenever 
velocity dispersion does not play a role.  The wavefunction description 
has the advantage that it can describe phenomena associated with 
velocity dispersion, such as free-streaming, and that it can be used
not only in the linear regime but also in the non-linear regime, after 
shell crossing.

We asked whether the Schr\"odinger-Poisson equations correctly describe 
axion dark matter.  The answer is {\it yes} on time scales short compared 
to the relaxation time scale $\tau$, and {\it no} on time scales long 
compared to $\tau$.  Whenever $\tau$ is shorter than the age of the universe 
$t$, axion dark matter is not correctly described by the Poisson-Schr\"odinger 
equations.  Indeed axions move towards a Bose-Einstein distribution on 
the time scale $\tau$ whereas the Schr\"odinger-Poisson equations would 
predict that they move towards a Boltzmann distribution.  Interactions, 
such as gravity or $\lambda \Phi^4$ interactions, play a dual role.  On 
the one hand the interaction influences the evolution of the classical 
field.  The solutions of the classical field equation, which is equivalent 
to the Schr\"odinger equation in the non-relativistic limit, describe the 
quantum states that the axions may occupy.  But the interaction has the 
additional role of causing the axions to jump between those states.  On 
time scales large compared to $\tau$, the distribution of quanta over the 
states described by the classical field changes completely.  It was shown 
in Refs.\cite{CABEC,therm} that, as a result of axion gravitational 
self-interactions, $\tau$ becomes and remains shorter than the age of 
the universe after the photon temperature has reached approximately 500 eV.  
Therefore the commonly made assumption that axion dark matter is adequately 
described by classical field equations at all times is incorrect.

In summary then, we have obtained the solutions of the Schr\"odinger-Poisson 
equations that describe the homogeneous expanding Friedmann matter-dominated
universe and density perturbations therein.  Each solution corresponds to 
a possible quantum-mechanical state of dark matter axions.  However the 
gravitational and other self-interactions of the axions cause them to jump 
between those states. The resulting dynamical evolution and observational 
implications of axion dark matter is the object of future work \cite{fut}.

%\acknowledgments (R3)
\begin{acknowledgments}

P.S. would like to thank Sankha Chakrabarty and Yaqi Han for useful 
discussions.  This work was supported in part by the U.S. Department 
of Energy under grant DE-FG02-97ER41209 at the University of Florida.

\end{acknowledgments}

%%%%%%%%%%%%%%%%%%%%%%%%%%%%%%%%%%%%%%%%%%%%%%%

\newpage

%\begin{references} (R3)

\end{document}